\documentclass[epjCONF]{svjour}
\usepackage{graphics}
\usepackage[varg]{txfonts} 
\usepackage[latin1]{inputenc}
\session-title{Third International Workshop on Compound Nuclear Reactions and Related Topics}
\begin{document}
\title{Nuclear Level Density of ${}^{161}$Dy in the Shell Model Monte Carlo Method}
\author{C. \"{O}zen \inst{1,2}\fnmsep\thanks{\email{cem.ozen@khas.edu.tr}} \and Y. Alhassid\inst{2} \and H. Nakada\inst{3}}
\institute{Faculty of Engineering and Natural Sciences, Kadir Has University, \\ Cibali 34083, Istanbul, Turkey \and Center for Theoretical Physics, Sloane Physics Laboratory,
Yale University, \\ New Haven,  CT 06520, USA
\and Department of Physics, Graduate School of Science,  Chiba University, Inage,\\  Chiba 263-8522, Japan}
\abstract{
We extend the shell-model Monte Carlo applications to the rare-earth region to include the odd-even nucleus ${}^{161}$Dy. The projection on an odd number of particles leads to a sign problem at low temperatures making it impractical to extract the ground-state energy in direct calculations. We use level counting data at low energies and neutron resonance data to extract the shell model ground-state energy to good precision.  We then calculate the level density of ${}^{161}$Dy and find it in very good agreement with the level density extracted from experimental data.
} 
\maketitle
\section{Introduction}
\label{intro}
The shell model Monte Carlo (SMMC)~\cite{LANG93,ALH94,KOO98,ALH01} is a powerful method for the calculation
of statistical properties of nuclei at finite temperature.
This approach has proven to be particularly useful in the calculation of nuclear level densities~\cite{NAK97,ORM97,ALH99,ALH07,OZE07}.
Reliable microscopic calculations of the level density often require the inclusion of correlations beyond the mean-field approximation.  In the SMMC approach such correlations are treated in the context of the
interacting shell model. SMMC calculations can be carried out in model spaces that are many orders of magnitude larger than model spaces that can be used in the conventional matrix diagonalization approach to the shell model.

Recently, the SMMC approach was extended to the rare-earth region and applied to a well-deformed even-even nucleus ${}^{162}$Dy~\cite{ALH08}.
Here we extend the application of SMMC in rare-earth nuclei to include the odd-even nucleus ${}^{161}$Dy. The projection on an odd number of particles leads to a sign problem (even for good-sign interactions), making it impractical to calculate thermal observables at low temperatures. Consequently, the ground-state energy of the nucleus cannot be extracted from direct SMMC calculations at very low temperatures. Here we use level counting data at low energies and neutron resonance data to determine the ground-state energy. We then
calculate the SMMC level density of ${}^{161}$Dy  and compare it with the level density extracted from available data.

\section{Choice of Model Space and Interaction}

We use the model space of Ref.~\cite{ALH08}, where the single-particle orbitals are $0g_{7/2}$, $1d_{5/2}$, $1d_{3/2}$, $2s_{1/2}$, $0h_{11/2}$ and $1f_{7/2}$ for
proton, and $0h_{11/2}$, $0h_{9/2}$, $1f_{7/2}$, $1f_{5/2}$,
$2p_{3/2}$, $2p_{1/2}$, $0i_{13/2}$, and $1g_{9/2}$ for neutrons.
The single-particle energies are determined to coincide with the spherical Woods-Saxon plus spin-orbit potential in the spherical Hartree-Fock approximation.
The effective interaction consists of monopole pairing and multipole-multipole terms~\cite{ALH08}
\begin{equation}
 - \!\! \sum_{\nu=p,n} g_\nu P^\dagger_\nu P_\nu
 - \!\! \sum_\lambda \chi_\lambda : (O_{\lambda;p} + O_{\lambda;n})\cdot
 (O_{\lambda;p} + O_{\lambda;n})\!: \;,
\end{equation}
where the pair creation operator $P^\dagger_\nu$ is given by
$P^\dagger_\nu = \sum_{nljm}(-)^{j+m+l}a^\dagger_{\alpha jm;\nu}
a^\dagger_{\alpha j-m;\nu}$
and the multipole operator $O_{\lambda;\nu}$ is given by $O_{\lambda;\nu}=\frac{1}{\sqrt{2\lambda+1}}
\sum_{ab}\langle j_a||\frac{dV_\mathrm{WS}}{dr}Y_\lambda||j_b\rangle
[a^\dagger_{\alpha j_a;\nu}\times \tilde{a}_{\alpha j_b;\nu}]^{(\lambda)}$
with $\tilde{a}_{jm}=(-)^{j+m}a_{j-m}$.  $:\,:$ denotes normal ordering.
The pairing strengths, $g_\nu= \gamma \bar{g}_\nu$,
where $\bar{g}_p=10.9/Z$ and $\bar{g}_n=10.9/N$ are parametrized
to reproduce the experimental odd-even mass differences
for nearby spherical nuclei in the number-projected BCS approximation~\cite{ALH08}. The quadrupole, octupole and hexadecupole interaction terms have strengths given by $\chi_\lambda= k_\lambda\chi$
for $\lambda=2,3,4$ respectively. $\chi$ is determined self-consistently~\cite{ALH96} and $k_\lambda$ are renormalization factors accounting for core polarization effects. Here we use the values for $\gamma$ and $k_\lambda$ as in the study of $^{162}$Dy~\cite{ALH08}.

\section{Ground-state Energy}

Because of the sign problem introduced by the projection on an odd number of particles, the thermal energy $E(\beta)$ of $^{161}$Dy can in practice be calculated only up to $\beta \sim 5.5$ MeV${}^{-1}$. The calculations for $\beta>2$ MeV${}^{-1}$ were carried out using the recently implemented stabilization routines~\cite{ALH08} to stabilize the canonical propagator. Since the discretization of $\beta$ introduces
systematic errors in $E(\beta)$, we calculated the thermal energy at any given $\beta$ for two values for the time slice ($\Delta\beta=1/32$ MeV${}^{-1}$ and
$\Delta\beta=1/64$ MeV${}^{-1}$) and then performed an extrapolation to $\Delta\beta=0$.  For $\beta\leq 3$  MeV${}^{-1}$, a linear extrapolation was found to be
suitable while for larger values of $\beta$, the dependence of $E(\beta)$ on $\Delta\beta$ is weaker and we took an average value.

\begin{figure}
\begin{center}
\vspace{0.6cm}
\resizebox{0.75\columnwidth}{!}{
  \includegraphics{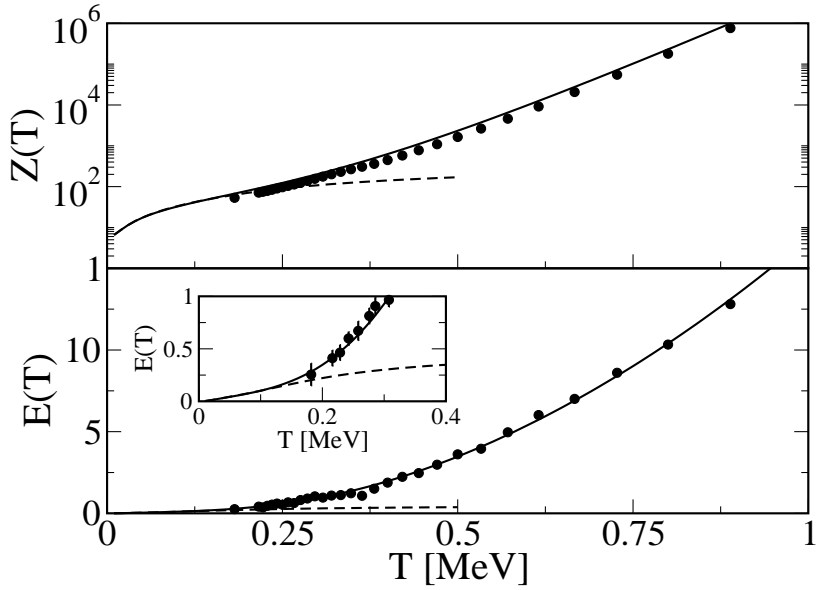} }
\caption{Partition function $Z$ (top panel) and thermal excitation energy $E^\ast$ (bottom panel) versus temperature $T$ for ${}^{161}$Dy.
The SMMC results (solid circles) are compared with the results deduced from experimentally known levels (dashed lines), and
from Eq.~(\ref{Eq:ZBBF}) (solid lines). 
The ground-state energy $E_0$ is obtained by a one-parameter fit of the SMMC thermal energy to the experimental thermal energy (solid curve). Inset: thermal energy $E$ versus temperature for $T< 0.4$ MeV.}
\label{fig:ZE}       
\end{center}
\end{figure}

To determine the ground-state energy $E_0$, we fitted (by a one-parameter fit) the SMMC thermal excitation energy $E^\ast(T)=E(T)-E_0$ to the experimental thermal excitation energy. The latter is calculated from $E^\ast(\beta)=-\partial \ln Z(\beta)/\partial\beta$, where $Z$ is the experimental partition function (see below).  The experimental partition function and the thermal excitation energy are shown, respectively, by solid lines in the top and bottom panels of Fig.~\ref{fig:ZE}. The solid circles are the SMMC results shown down to a temperature of $T\approx 0.18$ MeV, below which the statistical errors become too large because of the sign problem  (for comparison the lowest temperature calculated in the even-even nucleus $^{162}$Dy was $T=0.05$ MeV~\cite{ALH08}).

At sufficiently low temperatures the experimental partition function can be calculated from $Z(\beta)=\sum_{i} (2J_i+1) e^{-\beta E_i}$ where $E_i$ are the experimentally known energy levels of the nucleus (measured with respect to the ground-state energy). This partition function and the corresponding thermal excitation energy are shown by the dashed lines in Fig.~\ref{fig:ZE}.  However, since the level counting data is incomplete above a certain excitation energy, the average experimental thermal energy becomes saturated at temperatures above $T\sim 0.18$ MeV (see the inset in the lower panel of Fig.~\ref{fig:ZE}). Since SMMC data exist only above $T\sim 0.18$ MeV, a realistic estimate of the experimental thermal energy at higher temperatures is necessary to determine $E_0$ from the fit. We accomplish this by using an experimental partition function defined by
\begin{equation}
   Z(T)=\sum_{i}^{N} (2J_i+1) e^{-E_i/T} +\int_{E_{N}}^\infty d E \rho_{{BBF}}(E) e^{-E/T}\;,
   \label{Eq:ZBBF}
\end{equation}
where $E_N$ is an energy below which an essentially complete set of levels is known and $\rho_{BBF}(E)=\frac{\sqrt{\pi}}{12}a^{-1/4}(E-{\Delta})^{-5/4}e^{2\sqrt{a(E-{\Delta})}}$ is an experimentally determined level density parametrized by the back-shifted Bethe formula (BBF)~\cite{DILG73}. The single-particle level density parameter  $a$ and the backshift $\Delta$ are determined from level counting data at low energies and the $s$-wave neutron resonance data at the neutron resonance energy.
We find $a=18.564$ MeV$^{-1}$ and $\Delta= -0.615$ MeV.  The solid lines in 
Fig.~\ref{fig:ZE} are then calculated from 
Eq.~(\ref{Eq:ZBBF}). The number of the low-lying states $N$ (determining $E_N$) is chosen such that the two curves (solid and dashed) for the experimental partition function merge smoothly at a sufficiently low temperature. We found that $N=8$ is a reasonable choice.
Fitting the solid curve in the top panel of Fig.~\ref{fig:ZE} to the SMMC thermal excitation energy (solid circles), we found a ground-state energy of  $E_0=-363.920 \pm 0.020$ MeV.
We emphasize that the SMMC partition function and thermal energy fit very well the corresponding experimental quantities by adjusting only a single parameter $E_0$.

\section{Level Density}

\begin{figure}[h!]
\begin{center}
\vspace{0.6cm}
\resizebox{0.75\columnwidth}{!}{
  \includegraphics{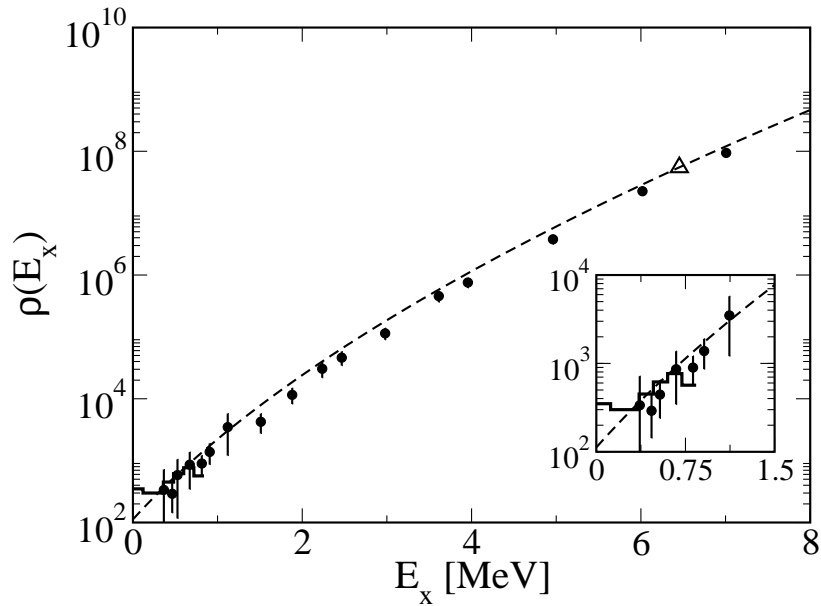} }
\caption{Level density of ${}^{161}$Dy. The SMMC results (solid circles) are compared with experimental results. The
histograms are from level counting data, the triangle is the neutron resonance data and the dashed line is the experimentally extracted BBF level density. Inset: level density for excitation energies $E_x<1.5$ MeV.}.
\label{fig:rho}       
\end{center}
\end{figure}

The level density is the inverse Laplace transform of the partition function. Its average is determined in the saddle-point approximation in terms of the canonical entropy and the heat capacity, which in turn are calculated from the average thermal energy $E(\beta)$~\cite{NAK97}.
The SMMC level density is shown by the solid circles in Fig.~\ref{fig:rho}. We compare it with the experimentally determined BBF level density (dashed line). The level counting data at low energies are shown by the histograms, and the neutron resonance data is shown by the triangle at $E_x=6.454$ MeV.
We find that at low excitation energies and close to the neutron resonance energy, the SMMC results agree quite well with the experimental level density.
At intermediate energies, the SMMC level density is slightly below the BBF level density. 

\section{Conclusions}

We have extended the SMMC calculations in the rare-earth region to include the odd-even nucleus, ${}^{161}$Dy. The sign problem introduced by the projection on the odd number of particles makes it impractical to calculate directly the ground-state energy. We have circumvented this problem by using the neutron resonance data and  level counting data at low energies. We have then calculated the level density and find it to be in good agreement with the experimental level density extracted from available data. The method outlined here requires sufficient level counting data at low energies and neutron resonance data.

This work is supported in part by the U.S. DOE grant
No. DE-FG-0291-ER-40608 and as Grant-in-Aid for Scientific Research (C), No. 22540266, by the MEXT, Japan.


\begin{thebibliography}{}
\bibitem{LANG93} G.~H.~Lang, C.~W.~Johnson, S.~E.~Koonin and W.~E.~Ornmand,
Phys.~Rev.~C \textbf{48} (1993) 1518.
\bibitem{ALH94} Y.~Alhassid, D.~J.~Dean, S.~E.~Koonin, G.~Lang, and W.~E.~Ormand,
Phys.~Rev.~Lett. \textbf{72} (1994) 613.
\bibitem{KOO98} S.~E.~Koonin, D.~J.~Dean, K.~Langanke, Phys.~Rep. \textbf{278} (1997) 1.
\bibitem{ALH01} Y.~Alhassid, Int. J. Mod. Phys. B {\bf 15} (2001) 1447.
\bibitem{NAK97} H.~Nakada and Y.~Alhassid, Phys.~Rev.~Lett. \textbf{79} (1997) 2939.
\bibitem{ORM97} W.~E.~Ormand, Phys.~Rev. {\bf C56}, (1997) R1678.    
\bibitem{ALH99} Y.~Alhassid, S.~Liu, and H~Nakada, Phys.~Rev.~Lett. \textbf{83} (1999) 4265.
\bibitem{ALH07} Y.~Alhassid, S.~Liu, and H~Nakada, Phys.~Rev.~Lett. \textbf{99} (2007) 162504.
\bibitem{OZE07} C.~\"{O}zen, K.~Langanke, G.~Martinez-Pinedo, and D.~J.~Dean, Phys. Rev. C \textbf{75} (2007) 064307.
\bibitem{ALH08} Y. Alhassid, L. Fang and H. Nakada, Phys.~Rev.~Lett. \textbf{101} (2008) 082501.
\bibitem{ALH96} Y. Alhassid, G.~F.~Bertsch, D.~J.~Dean, and S.~E.~Koonin, Phys. Rev. Lett. \textbf{77} (1996) 1444.
\bibitem{DILG73} W.~Dilg, W.~Schantl, H.~Vonach, M.~Uhl, Nucl.~Phys.~A \textbf{217} (1973) 269.
\end{thebibliography}
\end{document}